# Local field of spin-spin interactions in the nuclear spin system of *n*-GaAs

V. M. Litvyak and R. V. Cherbunin

*Spin Optics Laboratory, St. Petersburg State University, Ulyanovskaya 1, St. Petersburg 198504, Russia*

V. K. Kalevich[1,2]

[1]*Ioffe Institute, Politekhnicheskaya 26, St. Petersburg 194021, Russia*

[2]*Spin Optics Laboratory, St. Petersburg State University, Ulyanovskaya 1, St. Petersburg 198504, Russia*

K. V. Kavokin

*Spin Optics Laboratory, St. Petersburg State University, Ulyanovskaya 1, St. Petersburg 198504, Russia*

At low lattice temperatures the nuclear spins in a solid form a closed thermodynamic system that is well isolated from the lattice. Thermodynamic properties of the nuclear spin system are characterized by the local field of spin-spin interactions, which determines its heat capacity and the minimal achievable nuclear spin temperature in demagnetization experiments. We report the results of measurement of the local field for the nuclear spin system in GaAs, which is a model material for semiconductor spintronics. The choice of the structure, a weakly doped GaAs epitaxial layer with weak residual deformations, and of the measurement method, the adiabatic demagnetization of optically cooled nuclear spins, allowed us to refine the value of nuclear spin-spin local field, which turned out to be two times less than one previously obtained. Our experimental results are supported by calculations, which take into account dipole-dipole and indirect (pseudodipolar and "exchange") nuclear spin interactions as well as quadrupole splitting of nuclear spins in the vicinity of charged impurity centers.

## I. INTRODUCTION

In solids at low enough temperatures, the nuclear spins of the lattice very weakly interact with phonons, that leads to their effective thermal insulation. At the same time, the nuclear spin-spin interactions remain efficient. Under these conditions, the nuclear spin system (NSS) quickly comes to the state described by the density matrix with all the off-diagonal elements close to zero. The relations between diagonal elements are given by the Boltzmann factor with the effective nuclear spin temperature, which can differ from the lattice temperature by several orders of magnitude. As a result, the macroscopic characteristics of the NSS, and first of all its magnetization, obey the laws of thermodynamics. This fact is demonstrated in classic experiments by Pursell an Pound, Abragam with coauthors and others [1-4]. The most important parameter of the NSS is the local field, which determines the thermodynamic properties of the NSS. The local field characterizes the magnitude of nuclear spin-spin interactions and also their contribution to heat capacity and susceptibility of the NSS [5-6].

The above is indeed true for the NSS in semiconductors. Optical detection of electron and nuclear spin polarization, as well as the possibility of dynamic polarization of nuclear spins by optically oriented electron spins [6-7] allow one to investigate average value and fluctuations of nuclear magnetization in low magnetic fields, down to zero field. Under such conditions, the nuclear spin-spin interactions are most clearly manifested.

Gallium arsenide is the basic model material for studying the properties of the NSS in semiconductor structures. Arsenic and both isotopes of gallium (with atomic numbers 69 and 71) have nuclear spin moments equal to $I = 3/2$, so that all nuclei of the gallium arsenide lattice contribute to the nuclear magnetization. Most of the initial experiments on optical cooling of the NSS were carried out in bulk GaAs structures. The possibility of cooling the NSS in GaAs by optical pumping to temperatures of the order of several microkelvins was demonstrated [8-10]. It is also shown in those works that the cooled NSS absorbs the power of an external oscillating magnetic field at the resonance frequencies of nuclear isotopes. This makes it possible to obtain absorption spectra of the NSS in zero and external magnetic fields [10-13]. In particular, the absorption spectra in zero magnetic field reveal quadrupole effects, with magnitudes large enough to mask the manifestations of nuclear spin-spin interactions. Taking them into account is important in studying the thermodynamic properties of nuclear spins, especially the nuclear local field.

The local field was initially introduced as a characteristic of nuclear spin-spin interactions [3], so that the squared local field is proportional to mean squared energy of these interactions. However, it is known from resent works on cooling of the NSS in GaAs, that the value of local field is strongly affected by quadrupolar splitting of nuclear spin energy levels. Quadrupole effects due to crystal deformation increase the local field up to several gauss [9,10,12-14]. Taking this effect into account is very important for studying thermodynamic properties of the NSS in structures with quantum dots, quantum wells and microcavities [9,10,14].

The local field of spin-spin interactions in undeformed gallium arsenide was calculated in Ref. [15]: $B_{LSS} \approx 1.45$ G ($B_{LSS}^2 = 2.1 \pm 0.1$ G$^2$). In Ref. [15], and also in Refs. [9-14, 16], the contribution to the local field from quadrupole interactions was also considered. The mean squared energy of quadrupole interaction is characterized by the value $B_{LQ}^2$, which adds up to the squared spin-spin local field $B_{LSS}^2$, the squared effective local field being introduced: $B_{Leff}^2 = B_{LSS}^2 + B_{LQ}^2$. Such notations for contributions to the local field will be used in theoretical part of this work. The local field determined in experiments will be denoted as $B_L$ (including all possible contributions) throughout the text.

At the same time, Ref. [15] presents the experimental results on determining the local field by measuring the efficiency of nuclear dynamic polarization by electrons as a function of the external magnetic field in heavily doped and compensated *p*-GaAs. This way, the value of $\xi B_L^2 = 6.2 \pm 1.0$ G² was measured, where the factor $\xi = 2.2 \pm 0.3$, characterising relative contributions of various interactions to the nuclear spin dynamics, was calculated theoretically. The experimentally measured value of the local field was approximately 1.7 G. A slight excess of the experimentally determined value of the local field over the theoretically calculated value in Ref. [15] was qualitatively explained by the influence of the quadrupole interaction, which cannot be completely eliminated in a heavily doped material due to the electric fields of charged impurities, inevitably present there.

It should be noted, that the indirect method of determining the value of the nuclear local field by measuring the combination $\xi B_L^2$ in Ref. [15], and also the use of heavily doped and compensated *p*-GaAs, where quadrupole effects due to built-in electric fields are inevitable, leaves some uncertainly of the obtained value $B_L = 1.7$ G. These nuances of determining the local field of the NSS in GaAs in Ref. [15] inspired us to carry out direct measurements of the $B_L$, which are based on demagnetization experiments. Such experiments allow one to directly determine the value of $B_L$ as a thermodynamic characteristic of the NSS in GaAs.

For this purpose we performed the measurements of the local field in lightly doped *n*-type GaAs with the donor concentration $n_D = 2 \cdot 10^{15}$ cm⁻³. Experiments were done in an epitaxial layer of *n*-GaAs with low residual deformations, which made it possible to minimize the effect of the strain-induced quadrupole splitting on the magnitude of the measured local field. Since our measurements were carried out by an optical method, via the detection of the degree of photoluminescence circular polarization, we also took into account possible influence of the field created by optically pumped electron spins, $B_e$ (the Knight field) [6, Chapters 5 and 9]. We suppose that it can influence the determined value of the local field $B_L$, because the demagnetization process occurs in the total field, $\sqrt{B_L^2 + B_e^2}$. For this reason, we did some of our experiments under the conditions of elliptical optical pumping, in order to decrease the value of $B_e$ during the measurement process and to estimate the influence of the electron spin polarization on the measured local field. For realization of effective optical cooling of NSS in the case of elliptical pumping, our experiments were performed under the condition of different optical cooling times: 7, 20 and 27 seconds. As a result of increasing this time, measured magnitudes of local field $B_L$ decreased. We explain such a difference by influence of the electric charge of the donor center. Low impurity concentration in our sample made it possible to take into account the built-in electric fields around donor centers, bringing about a quadrupole contribution $B_{LQ}$ to the spin-spin local field $B_{LSS}$. Such quadrupole contribution adds up to the pure spin-spin local field. This effect manifests itself most strongly in the experiment under short optical cooling time (7 seconds). The quadrupole contribution ($B_{LQ}$), as well as that of the Knight field of spin-polarized electrons ($B_{Le}$), are taken into account in the framework of a numerical model, that includes nuclear spin diffusion and relaxation due to hyperfine interaction with localized electrons.

The spin-spin local field $B_{LSS}$ is formed by dipole-dipole, pseudodipolar and scalar (exchange) nuclear spin interactions. The minimum value of the local field, measured under the experimental conditions that minimize the influence of built-in electric fields and the Knight field (in the case of elliptical pumping, under optical cooling times 20 and 27 seconds), is supposed to be close to $B_{LSS}$. Actually, $B_L$ measured under these conditions is close to the minimum value of the local field of spin-spin interactions, $B_{LSS} \approx 0.8\,\text{G}$, corresponding to compensation of the dipole-dipole interaction of nearest neighbors by their pseudodipolar interaction, in the absence of scalar exchange interaction. As noted in Introduction, constants of indirect interaction under optical pumping [18-19] could differ from those known from NMR experiments [17]. Remarkably, this value is almost twice smaller than that reported in Ref. [15], though all these types of internuclear interactions were also taken into account in calculations in Ref. [15]

## II. SAMPLE AND EXPERIMENTAL METHOD

For the measurements of the local field, we chose a bulk *n*-GaAs:Si crystal with donor concentration $n_D = 2 \cdot 10^{15}$ cm$^{-3}$. The sample comprised a 20 microns thick active layer grown by liquid-phase epitaxy on a GaAs substrate. As mentioned in the Introduction, it is known from recent works [9-14, 16] that the value of local field in GaAs-based heterostructures is determined not only by spin-spin interactions, but also by the quadrupole interaction induced by strains due to lattice mismatch. Our previous studies of bulk *n*-GaAs crystals by nuclear spin warm-up spectroscopy showed that such structures can also contain residual deformations leading to quadrupole splitting of nuclear spin levels and increasing the local field [12,13,16]. This method is based on the warm-up of the optical cooled NSS by radiofrequency (RF) field, and allows one to measure absorption spectra of the NSS (an absorption spectrum being the rate of the NSS heating $1/T_{RF}(f)$ as a function of the frequency $f$) in zero external magnetic field. Spin-spin interactions lead to appearance of absorption lines at frequencies about 1-4 kHz, corresponding to the spin precession of various isotopes in the local magnetic fields of neighboring nuclei. At presence of quadrupole interaction exceeding the spin-spin interactions, the absorption spectrum contains two peaks (blue spectrum in Fig. 1). The high-frequency peak corresponds to the splitting of nuclear spin energy levels of isotopes $^{75}$As, $^{69}$Ga and $^{71}$Ga by quadrupole interactions, and its position can change depending on the value of quadrupole splitting [13].

For the *n*-GaAs crystal studied in this work, the absorption spectrum in zero magnetic field has only one absorption peak at the frequency of about 3.5 kHz (red spectrum in Fig. 1). This allows us to conclude that residual deformations in the crystal under study are insignificant, and to expect a minimum quadrupole contribution to the measured value of the local field (here the quadrupole contribution caused by deformation is implied; quadrupole contribution from the built-in electric fields around donor centers will be taken into account and will play an important role for explanation of our experimental results. However, the magnitude of such a quadrupole contribution is less than nuclear spin-spin interaction and did not appear as additional high-frequency peak).

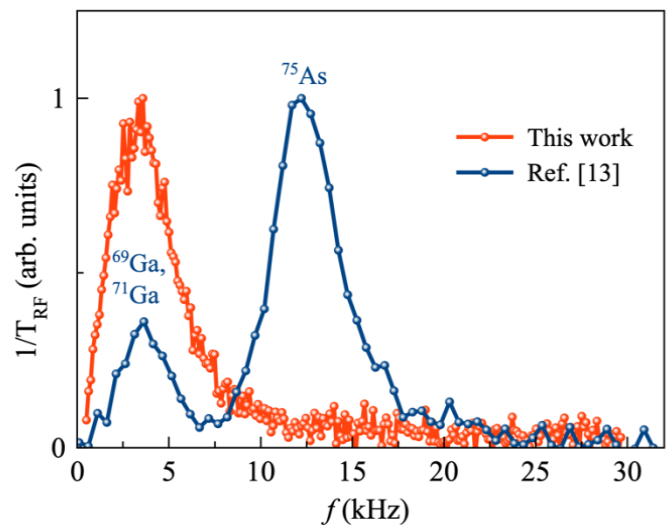

FIG. 1. Absorption spectra of RF power of an optically cooled NSS, measured in zero static external magnetic field for bulk layers of *n*-GaAs, differing from each other by the value of residual deformation. Red spectrum, which is measured in this work, has only one low-frequency absorption peak, that indicates insignificant residual deformation in the crystal under study. Blue spectrum, which was obtained in our previous work [13], has an additional high-frequency peak. This peak appears due to the nuclear quadrupole interaction, caused by residual deformation in the GaAs layer.

The local field measurement method we use is based on adiabatic demagnetization of the NSS, which has been preliminary cooled by optical pumping in a longitudinal magnetic field. According to the nuclear spin temperature theory [4], under adiabatic (slow on the spin-spin relaxation time scale, but fast on the spin-lattice relaxation time scale) variation of the external magnetic field, nuclear spin temperature changes according to the law [3-4]:

$$\frac{\theta_N(B)}{\theta_N(B_i)} = \sqrt{\frac{B_L^2 + B^2}{B_L^2 + B_i^2}}, \quad (1)$$

where $\theta_N$ is the nuclear spin temperature, $B_i$ is a magnetic field in which the NSS was cooled, $B$ is a magnetic field in which demagnetization occurs and $B_L$ is the nuclear local field.

The spin polarization of nuclei as a function of magnetic field, $p_N(B)$, is determined by the expression [7]:

$$p_N(B) = \frac{\hbar \langle \gamma_N \rangle B(I+1)}{3k_B \theta_N(B)}, \quad (2)$$

where $\langle \gamma_N \rangle = \frac{1}{2} \sum_k x_k \gamma_k$ is the nuclear gyromagnetic ratio averaged over all nuclear species (here $x_k$ and $\gamma_k$ are natural abundance and gyromagnetic ratio of $k$-th isotope, the factor ½ accounts for the fact that the primitive cell contains two atoms, As and Ga), $k_B$ is the Boltzmann constant, and $I = 3/2$ is the nuclear spin.

Nuclear spin polarization leads to appearance to the Overhauser field $B_N$ (nuclear field) acting on electrons, which is directly proportional to nuclear spin polarization [6, Chapter 5]:

$$B_N = b_N p_N, \quad (3)$$

where for GaAs $b_N \approx 5.3$ T [5].

Fitting of experimental curves $B_N(B)$ by Eqs. (1-3) allows one to determine the value of the local field $B_L$. Unlike the method used in Ref. [15], adiabatic demagnetization method does not require the determination of the factor $\xi$ and gives a direct estimate of the value of local field.

The experimental setup for measurements of the local field by demagnetization method is described in Ref. [13]. The sample is placed in a closed-cycle cryostat and cooled to 6.5 K. Laser diode radiation at the wavelength of 780 nm, passed through a linear polarizer and a quarter-wave plate, is focused on the sample surface. The rotation angle of the quarter-wave plate relative to the polarizer axes determines the degree and sign of circular polarization of the pump beam.

Photoluminescence (PL), passed through a photoelastic modulator (PEM) and a linear polarizer (GT), is focused on the spectrometer slit. The spectrometer passes the PL band at 817 nm, which is then focused on an avalanche photodiode (APD). Circular polarization of the PL is measured using a two-channel photon counter synchronized with the PEM.

For measurement of the nuclear field $B_N$ after adiabatic demagnetization of the optically cooled NSS, an experimental method, proposed in Ref. [11] and modified in Ref. [13] for measuring of absorption spectra, is used. First, the NSS is kept in the dark for 20 seconds at the presence of oscillating magnetic field (OMF) at the frequency of 3 kHz in order to erase the traces of its cooling during previous measurement cycles (thermalization).

At the second stage, the optical cooling is performed, during which polarized pumping and longitudinal magnetic field $B_Z = 150$ G are switched on. To reveal and eliminate the possible effect of the hyperfine field of optically oriented electrons (the Knight field $B_e$) on the dependence $B_N(B)$, experiments are performed at different absolute values of the degree of circular polarization $R$ of the pumping light (as noted above, at elliptical pumping, $R < 1$), and also at different signs of $R$. In this case, the sign of the circular polarization of light determined the sign of nuclear spin temperature. The value of inverse nuclear spin temperature $\beta = 1/\theta_N$, obtained as a result of optical pumping, is proportional to the product of the magnitude of the longitudinal magnetic field $B_Z$, in which optical cooling takes place, and z - projection of average electron spin, $S_Z$: $\beta \propto S_Z B_Z$ [6, Chapter 5]. The thermal equilibrium electron spin polarization under the conditions of our experiments is two orders of magnitude less than photoexcited one, and for this reason it does not play any role in the dynamic polarization of nuclear spins. Three series of measurements were done for positive ($\theta_N > 0$) and negative ($\theta_N < 0$) nuclear spin temperatures, differing in the absolute value of $R$, which was controlled by rotating of the quarter-wave plate to take values $R = 1, 0.87$ and $0.64$ (the corresponding rotation angles of quarter-wave plate are 45º, 30º and 20º). With decreasing $|R|$, electron spin polarization decreases too and, as a consequence, the optical cooling efficiency drops. In order to obtain the similar efficiency of nuclear optical cooling at the end of the second stage for all three experimental series, with decreasing of $|R|$, the optical cooling times are simultaneously increased. These times were 7, 20 and 27 seconds for $R = 1, 0.87$ and $0.64$, respectively. After pumping, the longitudinal magnetic field $B_Z = 150$ G is switched off within 20 ms, providing the adiabatic demagnetization of the NSS and achieving a lower absolute nuclear spin temperature $\theta_N(0) \approx \theta_N(B_Z) \cdot B_L / B_Z$. The absolute value of spin temperature after adiabatic demagnetization was $|\theta_N(0)| \approx 7 \cdot 10^{-5}$ K.

The last stage of the protocol is a measurement stage: an external transverse magnetic field $B_X$ is switched on, in the range from 0.2 to 9.5 G. The value of $B_N$ is measured by its effect on the degree of circular polarization of PL $\rho$, which is equal to z-projection of the average electron spin. Electron spins are depolarized in the total field $B_X + B_N$. At the beginning of this stage, the nuclear field $B_N$ is formed proportional to the nuclear spin polarization induced by the field $B_X$

$$B_N(B_X) = b_N p_N(B_X). \qquad (4)$$

The dependence of $B_N(B_X)$, obtained as a result of processing of the three-stage curves (the fitting process is described below in the section "Experimental results"), is an example of the classic demagnetization curve of the NSS, and the characteristic value of $B_X$, at which the dependence of $B_N(B_X)$ begins to approach a plateau, is determined by the value of the nuclear local field.

## III. EXPERIMENTAL RESULTS

Examples of three-stage curves for the measuring field $B_X = 0.67$ G are presented in Figs. 2 and 3. In Fig. 2, curves for positive and negative circular polarization of the pump ($R = +1$ and $R = -1$) are shown. The curve for $R = -1$, when nuclear spins are cooled to positive temperatures $\theta_N > 0$, has a characteristic maximum due to compensation of the field $B_X$ by the nuclear field $B_N$ (in GaAs in the case of positive nuclear spin temperature the nuclear field is antiparallel to the external magnetic field [6, Chapter 5 and Chapter 9]. As mentioned above, to reveal the possible effect of the Knight field on the measured value of local field, three types of experiments were performed, differing by the absolute value of the degree of circular polarization of the pumping light, $|R|$. In Fig. 3 (a) and (b) examples of curves for two degrees of the circular polarization of the pumping light ($R = 1$ and $R = 0.64$ respectively) for the case of $\theta_N < 0$ are shown. For the first case, the optical cooling time was equal to 7 seconds, and for the second case (elliptical pumping) this time was equal to 27 seconds.

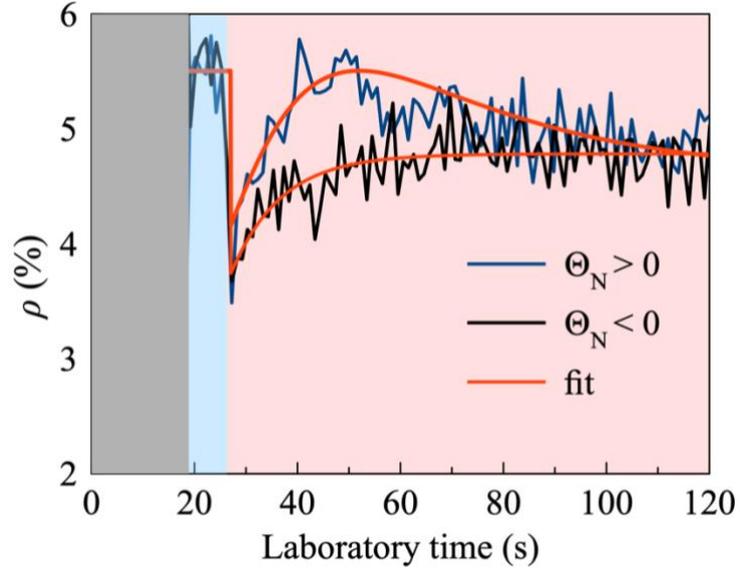

FIG. 2. Experimentally measured time dependences of PL circular polarization for positive (blue curve) and negative (black curve) nuclear spin temperatures. Red curves are obtained by fitting with Eq. (5).

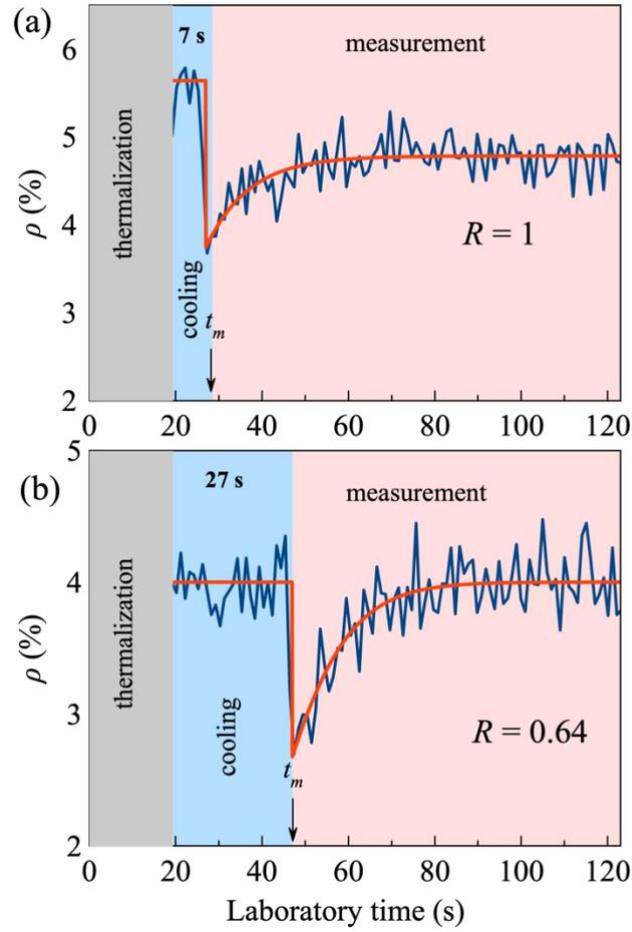

FIG. 3. Experimentally measured time dependences of PL circular polarization (blue curves): the NSS thermalization with lattice in the absence of optical pumping; optical cooling in longitudinal

magnetic field $B_Z = 150$ G during (a) 7 seconds for $R = 1$ and (b) 27 seconds for $R = 0.64$; the nuclear field $B_N$ measurement in presence of transverse magnetic field $B_X = 0.67$ G. Red curves are obtained by fitting with Eq. (5). These experimental curves correspond to the negative nuclear spin temperature $\theta_N < 0$.

In each of the experimental series, differing by value and sign of $R$, 14 measurements at different magnitudes of $B_X$ are done. The time dependence of the PL polarization $\rho(t)$ at the second ($t < t_m$) and third ($t > t_m$) stages for each of the measured curves is fitted by Eq. (5):

$$\rho(t) = \rho_0 \qquad (t < t_m)$$
$$\rho(t) = \rho_0 \frac{B_{1/2}^2}{B_{1/2}^2 + \left(B_x + b + (B_N - b)\exp\left(-\frac{t-t_m}{T_1}\right)\right)^2} \qquad (t > t_m), \qquad (5)$$

where $B_{1/2}$ is the HWHM of the Hanle curve, $t_m$ is the measurement start time, $T_1$ is the nuclear spin lattice relaxation time at presence of light, $b$ is the asymptotic value of nuclear field which develops due to dynamic polarization by optically oriented electron spins during the measurement stage. For the structure under study, $B_{1/2} \approx 40$ G and $T_1 \approx 10$ s. The value of $B_N$, formed at the time moment $t_m$ for a given magnitude of $B_X$, is used as a fitting parameter.

To determine the local field, we plot the dependences $B_N(B_X)$ (Fig. 4 (a), (b), dots), which are described by Eqs. (1-3) (Fig. 4 (a), (b), lines) with the value of local field $B_L$ as a fitting parameter.

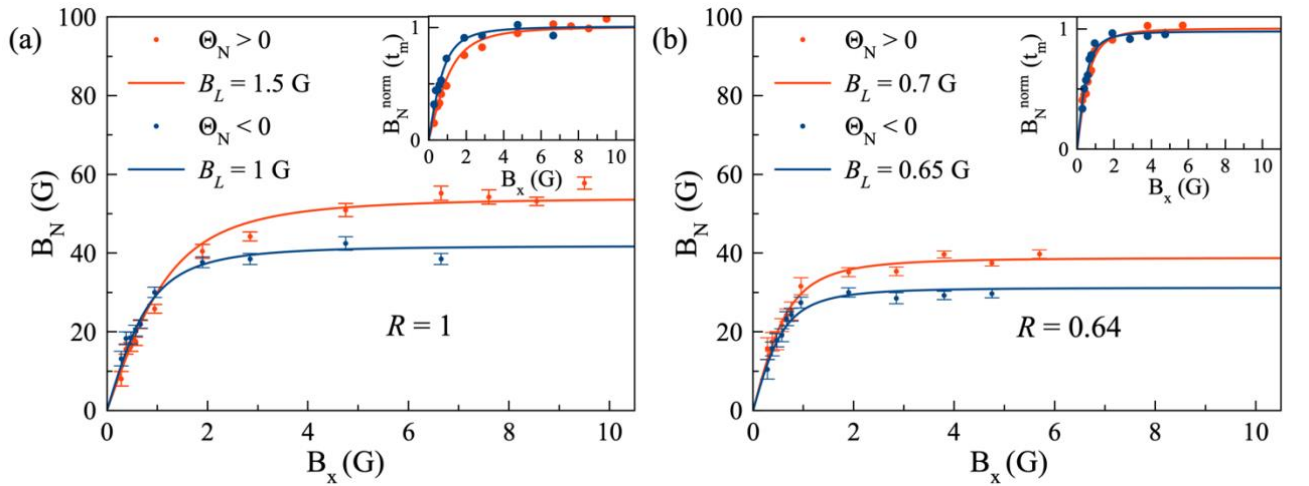

FIG. 4. Experimentally measured dependences of the nuclear field $B_N$ on the magnitude of transverse magnetic field $B_X$ (dots) for two degrees of the circular polarization of the pumping light: (a) $R = 1$ and (b) $R = 0.64$ for two signs of nuclear spin temperature (blue dots and lines

correspond to negative ($\theta_N < 0$), and red ones to positive ($\theta_N > 0$) nuclear spin temperatures). Solid red lines are results of fitting by Eqs. (1-3) with the value of local field (a) $B_L = 1.5$ G and (b) $B_L = 0.7$ G. Solid blue lines are results of fitting by Eqs. (1-3) with the value of local field (a) $B_L = 1$ G and (b) $B_L = 0.65$ G. Insets: dependences of $B_N(t_m)$ on $B_X$ normalized to the average saturation values of $B_N$ at large external field ($B \gg B_L$).

For the elliptical polarization of the pumping light $R = 0.64$, the local fields for two signs of spin temperature turned out to be close to each other: $B_L = 0.65$ G for $\theta_N < 0$ and $B_L = 0.7$ G for $\theta_N > 0$ (Fig. 4 (b)). For $R = 1$, the measured values of $B_L$ are significantly larger, and in addition, differ for two signs of spin temperature. This difference is not related to the fit error. This fact is shown in the insets of Fig. 4, which show the dependences $B_N(B_X)$. The value of $B_X$, at which $B_N(B_X)$ approaches the plateau, corresponds to the value of local field $B_L$. For |R| = 1, the dependence $B_N(B_X)$ approaches a constant level slower than in the case of |R| = 0.64. However, both local field values obtained from our direct demagnetization experiments on an undeformed lightly doped n-GaAs crystal differ from the calculated value of local field in Ref. [15], which is so far the generally accepted value for undeformed gallium arsenide.

The values of local field, obtained in our experiments, described above, are given in TABLE I (highlighted with gray background). In the last column of TABLE I the experimental value of $B_L$ from Ref. [15] is presented.

TABLE I. Experimentally measured values of the local field $B_L$ for negative $\theta_N < 0$ and positive $\theta_N > 0$ nuclear spin temperatures (4th and 5th columns) under different experimental conditions (1st, 2nd and 3rd columns). The last column shows experimentally measured local field $B_L$ from Ref. [15].

| Experimental conditions (this work) | | | $B_L$ (G) (this work) | | $B_L$ (G) (Exp.[a]) |
|---|---|---|---|---|---|
| R | quarter-wave plate rotation angle (deg) | $t_{pump}$ (s) | $\theta_N < 0$ | $\theta_N > 0$ | |
| 1 | 45 | 7 | 1 ± 0.1 | 1.5±0.1 | |
| 0.87 | 30 | 20 | 0.7±0.1 | 0.75±0.1 | 1.7 |
| 0.64 | 20 | 27 | 0.65±0.1 | 0.8±0.1 | |

[a] Reference [15].

## IV. THEORY

When the NSS in a semiconductor is studied by the optical orientation method, nuclear spins are dynamically polarized by electrons localized at impurity centers [6, Chapter 2]. As a consequence, the measured Overhauser fields are created by nuclei in the vicinity of donors, which are affected both by electron spins via the hyperfine interaction [5] and by charges of impurity centers via quadrupole interaction induced by electric fields [20]. Under these circumstances, the apparent local fields measured in optical experiments contain contributions from the Knight field and from quadrupole interaction.

The local field is defined as a rms of nuclear spin energy due to all interactions except the Zeeman interaction with the external magnetic field, divided by the rms of one of the Cartesian components of the nuclear magnetic moment in the absence of external field [4]. In our experimental geometry, where during the measurement stage the mean spin of photoexcited electrons is perpendicular to the external field, all the nuclear spin interactions, including hyperfine interaction with electrons, can be considered uncorrelated to each other and to the Zeeman interaction. Therefore, the squared effective local field $B_{Leff}^2$ can be presented as a sum of squared local fields of spin-spin interactions $B_{LSS}^2$, quadrupole $B_{LQ}^2$ and hyperfine interactions $B_{Le}^2$ [21]:

$$B_{Leff}^2(r) = B_{LSS}^2 + B_{LQ}^2(r) + B_{Le}^2(r). \qquad (6)$$

Out of these, only the local field of spin-spin interactions is the material parameter of the crystal; the others depend on the distance from the nearest donor and on the electron density, spin polarization during measurement and optical cooling times. As a result, the experimentally measured local field $B_L$ is an upper estimate of the "true" local field, i.e., one determined by spin-spin interactions, $B_{LSS}$. The measured local field would then exceed $B_{LSS}$ by certain value depending on the experimental conditions.

In order to estimate the apparent local field expected to be measured at certain experimental conditions, we first of all calculate $B_{LSS}$. Then, $B_{LQ}$ and $B_{Le}$ are calculated as functions of the distance from the donor center. Finally, the spatial distribution of nuclear polarization around the donor is evaluated, and the apparent value of local field is calculated for the conditions of our experiments.

## A. Calculation of the local field of spin-spin interactions in GaAs.

In accordance with the definition of the local field [4], its squared value is proportional to the trace of the square of the spin-spin Hamiltonian $\hat{H}_{SS}$:

$$B_{LSS}^2 = \frac{1}{N} \frac{3}{I(I+1)} \frac{1}{\hbar^2 \langle \gamma_N^2 \rangle} \text{Tr}\left\{ \left( \hat{H}_{SS} \right)^2 \right\}, \tag{7}$$

where $N$ is the number of nuclei in the ensemble, over the which the trace of $\hat{H}_{SS}$ is taken, $\langle \gamma_N^2 \rangle = \frac{1}{2} \sum_k x_k \gamma_k^2$ is the mean squared gyromagnetic ratio of nuclei, $k$ numerates the isotopes of Ga and As, $x_k$ is the abundance of $k$-th isotope.

For dipole-dipole interaction and other pair interactions, the trace of the square of the interaction must be calculated for the pair of interacting spins and averaged over all possible orientations of the pairs.

In addition to the dipole-dipole interaction, contribution to the local field can come from pseudodipolar and scalar ("exchange") interactions. These are indirect interactions mediated by valence electrons, and they are non-zero only for nearest neighbors. Manifestation of exchange interaction was investigated as broadening of NMR lines in III-V semiconductors (in particular, in GaAs) in [22 - 24]. The pseudodipolar interaction was mentioned in [25] also as a contribution to the second moment of NMR lines in GaAs. Later [26], also for a GaAs crystal, it was noted that the pseudodipolar interaction canceled a considerable part of the dipole-dipole interaction. The strenghts of pseudodipolar and exchange interactions are characterized by scalar constants: $B_{ij}$ and $A_{ij}$, respectively. The values of these constants were calculated and measured by NMR in GaAs crystals in early works [17, 24, 26]. As shown below, taking into account possible variation of $B_{ij}$ and $A_{ij}$ under optical pumping can explain low values of measured local fields.

The full Hamiltonian of spin-spin interactions equals [17]:

$$\hat{H}_{SS} = \frac{\hbar^2 \gamma_i \gamma_j}{r_{ij}^3} (1 + B_{ij}) \left[ \vec{I}_i \cdot \vec{I}_j - 3 \frac{(\vec{I}_i \cdot \vec{r}_{ij})(\vec{I}_j \cdot \vec{r}_{ij})}{r_{ij}^2} \right] + A_{ij} \frac{\hbar^2 \gamma_i \gamma_j}{r_{ij}^3}, \tag{8}$$

where the constants of pseudodipolar $B_{ij}$ and "exchange" $A_{ij}$ interactions are non-zero only for nearest neighbors. In GaAs, the nearest neighbor atoms are four atoms at the vertices of the tetrahedron, separated from the central atom by the distance $r_{nn} = a_0 \sqrt{3}/4$, where $a_0$ is the lattice constant.

For the pair of $i$-th and $j$-th spins:

$$\mathrm{Tr}\{H_{SS}^2\}_{ij} = \frac{(\hbar\gamma_i)^2(\hbar\gamma_j)^2}{r_{ij}^6}(1+B_{ij})^2 \times$$

$$\times \mathrm{Tr}\left\{\left[I_{ix}I_{jx}+I_{iy}I_{jy}+I_{iz}I_{jz}-3(X_{ij}I_{ix}+Y_{ij}I_{iy}+Z_{ij}I_{iz})(X_{ij}I_{jx}+Y_{ij}I_{jy}+Z_{ij}I_{jz})\right]^2\right\} + \quad (9)$$

$$+\frac{(\hbar\gamma_i)^2(\hbar\gamma_j)^2}{r_{ij}^6}A_{ij}^2\,\mathrm{Tr}\left\{\left[I_{ix}I_{jx}+I_{iy}I_{jy}+I_{iz}I_{jz}\right]^2\right\}$$

where $X_{ij} = \frac{\vec{r}_{ijx}}{r_{ij}}$, $Y_{ij} = \frac{\vec{r}_{ijy}}{r_{ij}}$, $Z_{ij} = \frac{\vec{r}_{ijz}}{r_{ij}}$.

Using properties of the trace: $\mathrm{Tr}\{I_{l\alpha}I_{k\beta}\} = \frac{I(I+1)}{3}\delta_{lk}\delta_{\alpha\beta}$, and also using the relation $X_{ij}^2 + Y_{ij}^2 + Z_{ij}^2 = 1$, after averaging over all possible positions of the of $i$-th and $j$-th spins, we can find:

$$\mathrm{Tr}\{H_{SS}^2\}_{ij} = 3\frac{(\hbar\gamma_i)^2(\hbar\gamma_j)^2}{r_{ij}^6}\left[\frac{I(I+1)}{3}\right]^2 \cdot \left[2(1+B_{ij})^2 + A_{ij}^2\right]. \quad (10)$$

In accordance with Eq. (7), the squared local field for the $j$-th spin equals:

$$B_{LSS}^2 = \frac{1}{2N}\frac{3}{I(I+1)\hbar^2\langle\gamma_N^2\rangle}\times 3\left[\frac{I(I+1)}{3}\right]^2$$

$$\times\left[\sum_{ij}\left(\frac{(\hbar\gamma_{As})^4}{r_{ij}^6}+\frac{\hbar^4\langle\gamma_{Ga}^2\rangle^2}{r_{ij}^6}\right)+\sum_{km}\left(\frac{\hbar^4\gamma_{As}^2\langle\gamma_{Ga}^2\rangle}{r_{km}^6}\right)\left[2(1+B_{km})^2+A_{km}^2\right]\right] = \quad (11)$$

$$= \frac{I(I+1)\hbar^2}{2\langle\gamma_N^2\rangle a_0^6}\cdot\left\{\left[\gamma_{As}^4+\langle\gamma_{Ga}^2\rangle^2\right]F_1 + 2\gamma_{As}^2\langle\gamma_{Ga}^2\rangle\left[F_2+(2B_{nn}+B_{nn}^2)F_3+\frac{F_3}{2}A_{nn}^2\right]\right\}$$

where indices $ij$ refer to the same sublattice (As and Ga), while indices $km$ refer to different sublattices. $B_{nn}$ and $A_{nn}$ are indirect interaction constants for nearest neighbors.

Thus, the calculation of the local field is reduced to the calculation of the dimensionless parameters: $F_1 = \sum_j \frac{a_0^6}{r_{0j}^6}$, $F_2 = \sum_k \frac{a_0^6}{r_{0k}^6}$ and $F_3 = \sum_{nn}\frac{a_0^6}{r_{nn}^6} = 4\frac{a_0^6}{r_{nn}^6} = 4\left(\frac{4}{\sqrt{3}}\right)^6 = 606.8$. Numerical summation gives $F_1 = 115.6$ and $F_2 = 660.6$.

The calculated value of the local field is obtained by substituting the GaAs parameters: $I = 3/2$, $\gamma_{As}/2\pi = 0.73\,\mathrm{kHz/G}$, $\gamma_{69Ga}/2\pi = 1.022\,\mathrm{kHz/G}$, $\gamma_{71Ga}/2\pi = 1.3\,\mathrm{kHz/G}$, $x_{69Ga} = 0.604$, $x_{71Ga} = 0.396$, $a_0 = 0.565$ nm into Eq. (11).

The calculated values of the local field of spin-spin interactions $B_{LSS}$ are given in TABLE II. In the case of pure dipole-dipole interaction ($B_{nn} = A_{nn} = 0$) we find that $B^2_{Ldd} \approx 2.46$ G² and $B_{Ldd} \approx 1.57$ G (1st column in TABLE II).

If we take the parameters of the pseudodipolar and "exchange" interactions from Ref. [17]: $B_{nn} \approx -0.6$ and $A_{nn} \approx 0.7$, we get the following values: $B^2_{Ldd} \approx 1.4$ G² and $B_{LSS} \approx 1.18$ G (2nd column in TABLE II). In the case, when pseudodipolar interaction totally compensates the magneto-dipole interaction for the nearest neighbors ($B_{nn} = -1$), while the exchange interaction is absent ($A_{nn} = 0$), the spin-spin local field equals $B_{LSS} \approx 0.8$ G (3rd column in TABLE II). The last case reflect the theoretical minimum value of spin-spin local field for GaAs and, as we will show below, this minimum value of spin-spin local field is close to our experimental results obtained for the degree of circular polarization of the pumping light $R = 0.87$ and 0.64 (under elliptical optical pumping). At the same time, the local field $B_{LSS} \approx 1.18$ G is close to the experimental result obtained for the degree of circular polarization of the pumping light $R = 1$: $B_L \approx 1$ G (see 4th column in TABLE I), but still exceeds it. This fact suggests that the measured local field, in addition to spin-spin, pseudodipolar and "exchange" interactions, can be influenced by the electric field gradients (EFG) created by the electric fields of localized electrons and/or the Knight field of photoexcited electrons. These effects will be estimated in the next section.

TABLE II. The calculated value of the local field for three types of spin-spin interactions (this work).

| Only dipole-dipole interaction $B_{nn} = A_{nn} = 0$ $B_{Ldd}$ (G) | Dipole-dipole + indirect interactions $B_{nn} \approx -0.6$ and $A_{nn} \approx 0.7$ $B_{LSS}$ (G) | Dipole-dipole + indirect interactions $B_{nn} = -1$ and $A_{nn} = 0$ $B_{LSS}$ (G) |
|---|---|---|
| 1.57 | 1.18 | 0.8 |

The values of the local field obtained in our calculations taking into account pseudodipolar and "exchange" interactions for GaAs for both cases, presented in TABLE II, turned out to be significantly lower than the value obtained in Ref. [15] taking into account the same interactions ($B_{LSS} \approx 1.45$ G, $B^2_{LSS} \approx 2.1$ G²). Possible reasons for this discrepancy will be discussed below.

## B. Calculation of the local field as a function of the distance to the donor center.

To explain the difference in the values of $B_L$ for different degrees of circular polarization of the pumping light (we will analyze experimental results only for the negative nuclear spin temperature: $\theta_N < 0$, see TABLE I, 4th column), we calculated the effect of the contribution of electric field created by the electron localized on the donor center [20], and also of the Knight field of this electron on the effective local field depending on the distance to the donor.

The electric field around a donor occupied by an electron is radial and, according to the Gauss theorem, at the distance $r$ from the center it equals to the Coulomb field created by the total electron charge $e + q_e(r)$ inside the sphere of the radius $r$:

$$E_D(r) = \frac{e + q_e(r)}{\varepsilon r^2}, \qquad (12)$$

where $\varepsilon$ is the dielectric constant,

$$q_e(r) = -e \int_0^r 4\pi \Psi_e^2(r') r'^2 dr' = -e \left[ 1 - \left( 1 + \frac{2r}{a_B} + \frac{1}{2}\left(\frac{2r}{a_B}\right)^2 \right) \exp\left(-\frac{2r}{a_B}\right) \right], \qquad (13)$$

and $\Psi_e(r) = \frac{1}{\sqrt{\pi a_B^3}} \exp\left(-\frac{r}{a_B}\right)$ is the wave function of the donor-bound electron.

In GaAs, the absence of the inversion symmetry results in a contribution to the local electric field gradient (EFG), which is linear in macroscopic electric fields (analogous to the piezo effect):

$$\frac{\partial^2 V}{\partial x_j \partial x_k} = R_{14} \nu_{jk,i} E_i, \qquad (14)$$

where $R_{14}$ is the tensor component relating EFG with homogeneous electric field, oriented along the z axis, $\nu_{jk,i} = \nu_{kj,i} = 1$ if $i \neq j \neq k \neq i$, and 0 otherwise.

The Hamiltonian of quadrupolar interaction is [5]:

$$\hat{H}_Q = \frac{eQ}{6I(2I-1)} \sum_{j,k} \left( \frac{\partial^2 V}{\partial x_j \partial x_k} \right) \left\{ \frac{3}{2}(I_j I_k + I_k I_j) - \delta_{jk} I(I+1) \right\}. \qquad (15)$$

By definition, quadrupole contribution to the squared local field is [21]:

$$B_{LQ}^2 = \frac{3}{I(I+1)} \frac{1}{\hbar^2 \langle \gamma_N^2 \rangle} \mathrm{Tr}\left\{\left(\hat{H}_Q\right)^2\right\}. \tag{16}$$

It follows from Eqs. (14-16) that $B_{LQ}^2$ is a quadratic function of the electric field components. Since in the cubic crystal the only second-order invariant is a scalar, $B_{LQ}^2$ doesn't depend on the direction of electric field. In order to calculate $B_{LQ}^2$, it is sufficient to find its value for the case of an electric field directed along the crystal axis [001], which we denote as z:

$$\begin{aligned} B_{LQ}^2 &= \frac{3}{4I^3(I+1)(2I-1)^2} \frac{e^2 \langle Q^2 \rangle R_{14}^2 E^2}{\hbar^2 \langle \gamma_N^2 \rangle} \mathrm{Tr}\left\{\left(I_X I_Y + I_Y I_X\right)^2\right\} = \\ &= \frac{3}{4I^3(I+1)(2I-1)^2} \frac{e^2 \langle Q^2 \rangle R_{14}^2 E^2}{\hbar^2 \langle \gamma_N^2 \rangle} \frac{I(I+1)}{15} \left\{4I(I+1)-3\right\} = \\ &= \frac{4I(I+1)-3}{20 I^2 (2I-1)^2} \frac{e^2 \langle Q^2 R_{14}^2 \rangle E^2}{\hbar^2 \langle \gamma_N^2 \rangle} \end{aligned} \tag{17}$$

The Knight field of localized electrons in semiconductors [6, Chapter 5] is proportional to squared wave function of the electron: $B_e(r) = \frac{A v_0 \langle S_e \rangle}{\gamma_N} |\Psi_e(r)|^2$, where $v_0$ is the primitive cell volume, and $\langle S_e \rangle$ is the mean electron spin. Taking into account also photoexcited electrons, we arrive to the following expression of the electron contribution to the squared local field:

$$B_{Le}^2(r) = \frac{\langle A^2 \rangle v_0^2 \langle S_e \rangle^2}{\langle \gamma_N^2 \rangle} \left[\frac{1}{\pi a_B^3} \exp(-2r/a_B) + n_c\right]^2, \tag{18}$$

where $n_c$ is the density of photoexcited electrons in the conduction band. For our experimental conditions, we estimated $n_c \approx 6 \cdot 10^{15}$ cm$^{-3}$.

In Fig. 5, the effective local field $B_{Leff}(r) = \sqrt{B_{LSS}^2 + B_{LQ}^2(r) + B_{Le}^2(r)}$ as a function of the distance to the donor, and also contributions to this local field from spin-spin $B_{LSS}$, quadrupole $B_{LQ}$ and hyperfine (the Knight field contribution $B_{Le}$) interactions are shown. The Knight field contribution is calculated for the case of circularly polarized pumping, when the mean electron spin, numerically equal to the PL circular polarization degree, was 0.06 (corresponding to the electron spin polarization 12%).

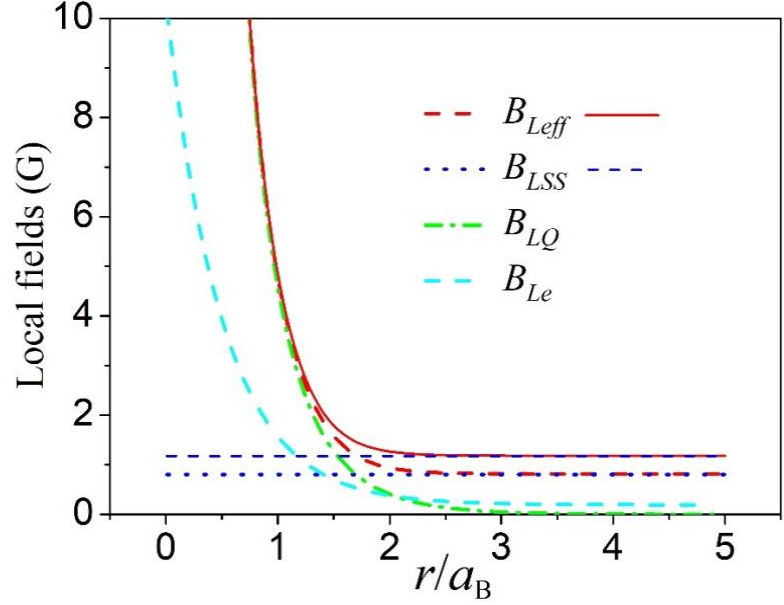

FIG. 5. Different contributions to the effective local field $B_{Leff}$ vs the distance to the donor center $r$. Blue dotted and dashed lines are local fields of nuclear spin-spin interactions $B_{LSS}$ for the case of different values of indirect constants: $B_{nn} = -1$, $A_{nn} = 0$ and $B_{nn} = -0.6$, $A_{nn} = 0.7$, respectively, light-blue dashed line is a contribution from the Knight field $B_{Le}$ at the electron spin polarization of 12% (in the experiment with the circularly polarized pumping); green dash-dotted curve is a contribution from the quadrupole interaction induced by electric fields around the donor, $B_{LQ}$. Solid and dashed red lines show the effective local fields $B_{Leff}$ with all these contributions in the case of $B_{nn} = -0.6$, $A_{nn} = 0.7$ and $B_{nn} = -1$, $A_{nn} = 0$, respectively.

The calculated curves presented in Fig. 5 demonstrate a sharp dependence of the effective local field, experienced by nuclear spins, on their distance from the nearest donor. Under these conditions, the local fields measured in the experiments should depend on the spatial distribution of the nuclear polarization, which is formed by the moment of measurement.

## C. Calculation of the spatial distribution of the nuclear spin polarization around the donor center.

In this section, the calculation of the spatial and temporal dependences of the nuclear spin polarization around the donor is presented. Dynamic polarization and relaxation of nuclear spins due to hyperfine interaction and nuclear spin diffusion are taken into account. The following approximations are used in the calculations: 1) the influence of other donors is not taken into account, i.e., it is assumed that the distance to the nearest donor is much greater than the nuclear spin diffusion length; 2) the filling number of the donor with electrons is taken equal to one.

In spherical coordinates, the nuclear spin polarization as a function of the distance to the donor center and time, $p_N(r,t)$, is described by the diffusion equation [27]:

$$\frac{\partial p_N(r,t)}{\partial t} = D \cdot \frac{1}{r^2} \frac{\partial}{\partial r}\left(r^2 \frac{\partial p_N(r,t)}{\partial r}\right) - W(r) p_N(r,t) + G(r), \tag{19}$$

where $D$ is a nuclear spin diffusion coefficient, $W(r)$ is the nuclear spin-lattice relaxation rate, and $G(r)$ is the rate of dynamic nuclear spin polarization during the pump. The relaxation and dynamic polarization occur due to the hyperfine interaction with the electron localized on the donor center. In this case, the value of $G(r)$ equals:

$$G(r) = \frac{I(I+1)}{s(s+1)} W(r) p_e, \tag{20}$$

where $s = \frac{1}{2}$ is the electron spin, $I = \frac{3}{2}$ is the nuclear spin, $p_e$ is the electron spin polarization [7].

The rate of nuclear spin-lattice relaxation equals:

$$W(r) = \frac{2}{3} s(s+1)\langle A^2\rangle \hbar^{-2} \left[v_0 |\Psi_e|^2\right]^2 \tau_c, \tag{21}$$

where $\langle A^2\rangle$ is the mean squared hyperfine interaction constant, $v_0$ is the primitive cell volume, $|\Psi_e|^2 = \frac{1}{\pi a_B^3} \exp(-2r/a_B)$ is the squared modulus of the localized electron wave function, and $\tau_c$ is the correlation time of the localized electron [6 - 7].

The substitution $f(r,t) = \frac{1}{r} F(r,t)$ converts the Eq. (19) into the following form:

$$\frac{\partial F(r,t)}{\partial t} = D \frac{\partial^2 F(r,t)}{\partial r} - W(r) F(r,t) + r G(r). \tag{22}$$

Equation (22) was solved numerically for the stages of pumping and measurement, besides zero polarization was taken as the initial condition at the stage of pumping: $F(r,t) = 0$ at $0 < r < 10 a_B$.

The result of the solution was the function $F_p(r,t)$, giving the polarization after pumping: $p_N(r,t_p) = \frac{1}{r} F_p(r,t_p)$. The polarization after adiabatic demagnetization was then obtained as $p_{NA}(r) = \frac{B_m}{\sqrt{B_{Leff}^2 + B_m^2}} p_N(r,t_p)$. For the calculation of the nuclear spin polarization at the time of measurement $t_m$, the Eq. (22) was solved again, with the function

$F_{NA}(r) = rp_{NA}(r) = \frac{B_m}{\sqrt{B_{Leff}^2 + B_m^2}} F_p(r, t_p)$ as the initial condition. The result of this solution was the function $F_m(r, t_m, B_m)$. It was then used to calculate the spatial dependence of the nuclear polarization at the moment of measurement:

$$p_N(r, t_m, B_m) = \frac{1}{r} F_m(r, t_m, B_m). \tag{23}$$

This polarization was then averaged with the effective electron density:

$$\rho_e(r) = \lambda \Psi_e^2(r) + n_c, \tag{24}$$

where $n_c$ is the concentration of photoexcited electrons in the conduction band, and $\lambda$ is the factor, which determines the relative contribution of localized electrons to the mean electron spin that manifests itself in the PL polarization.

As a result of averaging, the value of the experimentally measured Overhauser field is obtained:

$$B_{N\exp}(t_m, B_m) = \frac{\int_0^{n_D^{-1/3}} p_N(r, t_m, B_m) \rho_e(r) r^2 dr}{\int_0^{n_D^{-1/3}} \rho_e(r) r^2 dr}. \tag{25}$$

The results of the calculation of nuclear spin polarization in a high magnetic field ($B_m \gg B_{Leff}$) 0.5 second after the end of optical pumping (which corresponds to the beginning of the measurement stage in the experiment) are shown in Fig. 6 for two optical cooling times, 27 seconds (red curve, $R = 0.64$) and 7 seconds (blue curve, $R = 1$). Further, these numerical results were used for calculation of the Overhauser field and of the local field, measured within our experimental protocol.

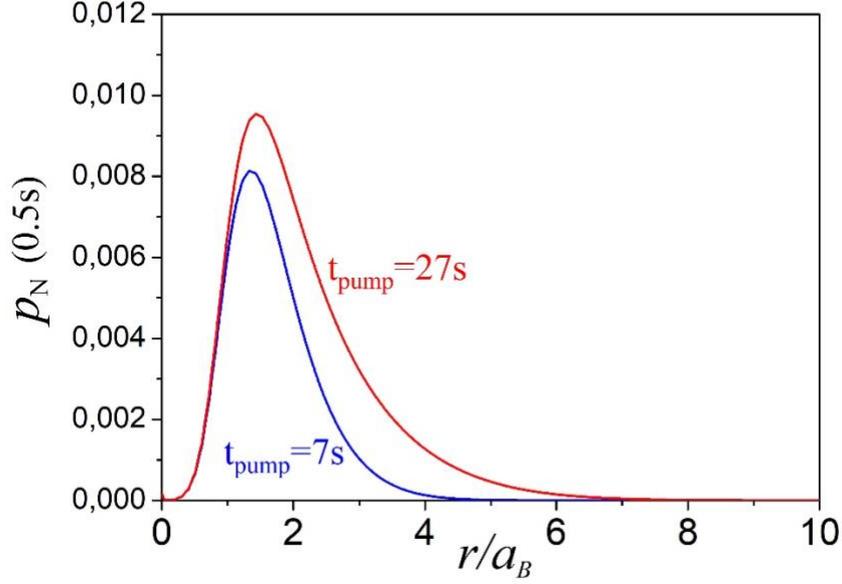

FIG. 6. The results of numerical calculation of spatially inhomogenious nuclear polarization in the vicinity of the donor center after 0.5 second from the end of optical pumping for two optical cooling times, 27 seconds (red curve, $R = 0.64$) and 7 seconds (blue curve, $R = 1$).

### D. Estimation of the experimentally measured local field $B_L$.

As follows from Eqs. (1-3), the local field can be expressed in terms of the ratio of nuclear polarizations (and, consequently, nuclear fields) in strong ($B_{m1} \gg B_L$) and weak ($B_{m2} \ll B_L$) measuring magnetic fields:

$$B_L = \frac{p_N(t_m, B_{m1})}{p_N(t_m, B_{m2})} B_{m2} = \frac{B_N(t_m, B_{m1})}{B_N(t_m, B_{m2})} B_{m2} \ . \tag{26}$$

This relation was used for calculation of the apparent local field observed in the experiment. The nuclear fields manifested in experiment were calculated using Eq. (25), by averaging the nuclear polarization with the electron density over the volume per one donor.

## V. DISCUSSION

The results of measurements of the local field at different degrees of circular polarization of the pumping light $R$ and different signs of the nuclear spin temperature are summarized in TABLE I. For comparison with theory, the experimental values of local field $B_L$ only for negative nuclear spin temperature $\theta_N < 0$ will be used. Explanation of difference between values of local field for $\theta_N < 0$ and $\theta_N > 0$ is outside the scope of this paper. We assume that this discrepancy may be due to the different shapes of experimental curves for $\theta_N < 0$ and $\theta_N > 0$ (see Fig. 2), which makes their fitting less precise for $\theta_N > 0$. But we do not exclude a physical cause for the observed difference. Experimentally measured and calculated values of the effective local field, which we use for discussion and comparison with each other are summarized in TABLE III.

TABLE III. Experimental conditions (1st and 2nd column) of measuring local fields $B_L$ and corresponding experimental results (3rd column). Calculated values of the effective local field $B_{Leff}$ under conditions corresponding to our experiments are summarized in two last columns.

| Experimental conditions | | Experimental results | Theory | |
|---|---|---|---|---|
| $R$ ($\theta_N < 0$) | $t_{pump}$ (s) | $B_L$ (G) | $B_{Leff}$ (G) ($\lambda=0$) | $B_{Leff}$ (G) ($\lambda=1$) |
| 1 | 7 | 1±0.1 | 1.07 | 1.38 |
| 0.87 | 20 | 0.7±0.1 | 0.91 | 1.12 |
| 0.64 | 27 | 0.65±0.1 | 0.88 | 1.07 |

It should be noted that a reasonably good agreement between local fields measured in our experiments and those calculated without taking into account the quadrupole interaction caused by crystal deformations confirms our initial assumption about a low level of residual deformation in lightly doped *n*-GaAs crystal under study.

First of all, we should discuss the origins of the difference between theoretical values of the local field obtained in Ref. [15] and in our work. Unfortunately, in Ref. [15], the details of the calculation are omitted, and the value of the dipole-dipole contribution to the local field (without taking into account indirect interactions) is not given separately. The parameters of indirect interactions in Ref. [15] are taken from Ref. [26]. Though we used the refined parameters from the later Ref. [17], where a somewhat larger absolute value of the pseudodipolar interaction parameter $B_{nn}$ is given (see 2nd column in TABLE II), the resulted difference is not big. The more probable reason is erroneous interpretation by Paget et al. of the numerical data from Ref. [26]. Indeed, in Ref. [15], with reference to Ref. [26], the value of $(1+B_{nn})^2 = 1.64 \pm 0.13$ is given. This would mean, that pseudodipolar interaction adds up to the magnetic dipole-dipole interaction and increases its contribution to the local field. At the same time, as it follows from the data presented in Table III in Ref. [26], the value of $(1+B_{nn})^2$ is in fact less than 1 and takes values from 0.22 to 0.49 for different GaAs samples studied there, i.e., the pseudodipolar interaction weakens the contribution of the magnetic dipolar interaction to the local field.

However, our experimental values of local field cannot be describe by using indirect constant from Ref. [26]. If we used these one (see 2$^{nd}$ column in TABLE II), calculated spin-spin local field ($B_{LSS} = 1.18$ G) exceed all experimental values (see 4$^{th}$ column in TABLE I). We tentatively attribute the decrease of the local field to changing the magnitude of indirect pseudodipolar interaction due to optical pumping, that can take place according to [18-19]. Our experimental results for $R = 0.87$ and $0.64$ ($B_L \approx 0.7$ G and $0.65$ G) are close to the theoretical value $B_{LSS} \approx 0.8$ G, obtained by taking $B_{nm} = -1$ and $A_{nm} = 0$, that is assuming total compensation of the magneto-dipole interaction between nearest neighbors by indirect interactions. Further experiments under the conditions excluding optical pumping during the measurement stage (e.g. by using spin noise spectroscopy [14]) are needed to clarify this issue.

As for the result for $R = 1$ ($B_L \approx 1$ G), we attribute the increase of the local field to the increase in the quadrupole contribution $B_{LQ}$ from the electric field of the donor center. The contribution of the Knight field of optically oriented electrons is insignificant, as can be seen from Fig. 5 and from a comparison of the calculated values of the effective local field for different experimental conditions (see TABLE III): the local field mainly depends on the optical cooling time and, to a small extent, on the electron polarization. Figure 6 shows that, for the two times of optical cooling, the nuclear polarization reaches its maxima at the distance from the donor center of about $3a_B$. However, for the optical cooling time equal to 27 seconds, spin diffusion leads to the propagation of the nuclear polarization up to the distance of about $8a_B$, while in 7 seconds the nuclear polarization propagates only to a distance of about $5a_B$. This leads to different contributions to the local field from the electric field of the donor center.

At the measurement stage, the Overhauser field acts on both localized and photoexcited free electrons. The measurement of PL polarization occurs at the photon energy close to the energy of free exciton, so that the polarization of delocalized electrons (bound into excitons or free) is directly measured. An opinion was expressed in the literature, based on theoretical estimations [28-29] and experiments on optically detected nuclear magnetic resonance [28, 30], that free and localized electrons form a common spin reservoir as a result of the fast exchange scattering. However, our calculations under such assumptions (formalized by the choice of the parameter $\lambda$ in Eq. (24) equal to 1) give values of the apparent local field that are overestimated compared to the experimental results. The best agreement with experiment is obtained in the model that assumes the detection of the nuclear polarization only through free electrons ($\lambda = 0$). This means that actually the spin exchange between localized and free electrons may be much less efficient at least under conditions of our experiment.

Finally, the difference between the local field values obtained experimentally for the positive and negative nuclear spin temperatures under circularly polarized pumping cannot be explained by our theory. As it exceeds the error of our measurements, we have to assume that it is due to some real effects that are not taken into account in our theory. Since the sign of the nuclear spin temperature is given by mutual orientation of the external magnetic field and the mean spin of photoexcited electrons during the optical pumping, we can assume that mutual orientation of the external magnetic field and spatially inhomogeneous Knight field of localized electrons during pumping can affect spatial distribution of the nuclear polarization. Verification of this assumption requires additional experimental and theoretical studies and is beyond the scope of this work.

## VI. CONCLUSION

In this work, the local field of the NSS was determined in a lightly doped epitaxial layer of *n*-GaAs with low level of residual deformation. Experiments on nuclear spins demagnetization, from which the value of local field was determined, were performed for three different degrees of circular polarization of pumping light and for two signs of the nuclear spin temperature.

By comparison of the experimental data and the results of theoretical modelling, we found that the local field measured in optical orientation experiments is determined not only by spin-spin interactions of nuclei, but also by spatially dependent agents: electric fields of donor centers and Knight fields of spin-polarized electrons. Among those, the electric fields, which induce quadrupole splitting of nuclear spin levels, are of the highest importance. As a result, the effective local field becomes spatially dependent on the scale of few donor Bohr radii. The results of experiments aimed at measuring the local field depend on averaging the effective local field over both spatial distribution of nuclear spin polarization and the density of photoexcited electrons. This fact leads to the dependence of the apparent local field on experimental conditions, such as pumping duration and ellipticity of the pump light.

A theoretical calculation of the local filed of spin-spin interactions was carried out, taking into account dipole-dipole, pseudodipolar and exchange nuclear spin interactions. The measured local fields under different conditions appear systematically lower than theoretical predictions. This points out to possible compensation of the magneto-dipole interaction of nearest neighbors by indirect interactions via electron states, which can be enhanced by optical pumping during measurement. The determined value of the local field of spin-spin interactions of approximately 0.7 G is two times less than the value of 1.45 G, obtained in Ref. [15] and commonly used in the literature.

## ACKNOWLEGMENTS

Experimental part of this work was supported by Russian Science Foundation, grant number 22-42-09020. The authors also acknowledge Saint-Petersburg State University for research grant number 94030557 for financial support of theoretical parts.